\documentclass[conference]{IEEEtran}
\usepackage{amsmath,amssymb,amsfonts}
\usepackage{algorithmic}
\usepackage{graphicx}
\usepackage{textcomp}
\usepackage{url}
\usepackage{siunitx}

\begin{document}


\title{Comparing Two Generations of Embedded GPUs Running a Feature Detection Algorithm}

\author{\IEEEauthorblockN{Max~Danielsson, H{\aa}kan~Grahn, and Thomas~Sievert}
\IEEEauthorblockA{
\textit{Blekinge Institute of Technology}\\
SE-371 79 Karlskrona, Sweden \\
max@autious.net, \{hakan.grahn,thomas.sievert\}@bth.se}
\and
\IEEEauthorblockN{Jim~Rasmusson}
\IEEEauthorblockA{
\textit{Sony Mobile Communications AB}\\
SE-221 88 Lund, Sweden \\
jim.rasmusson@sony.com}
}

\maketitle

\begin{abstract}
Graphics processing units (GPUs) in embedded mobile platforms are reaching performance
levels where they may be useful for computer vision applications. We compare two generations of embedded
GPUs for mobile devices when running a state-of-the-art feature detection algorithm,
i.e., Harris-Hessian/FREAK. We compare architectural differences, execution time,
temperature, and frequency on Sony Xperia Z3 and Sony Xperia XZ mobile devices.
Our results indicate that the performance soon is sufficient for real-time feature
detection, the GPUs have no temperature problems, and support for large work-groups
is important.
\end{abstract}




\section{Introduction}
\label{sec:introduction}
Today's cellphones have very powerful CPUs and embedded graphics processing units (GPUs) built into them.
For example, the Sony Xperia Z3~\cite{Z3} has a 2.5 GHz quadcore CPU and a 128 core Adreno 330 GPU.
This enables performance-demanding applications to migrate from desktop to mobile platforms.

Digital images play a large role in how we communicate with each other. As contemporary
cellphones are equipped with high-resolution digital cameras, the need for advanced and
powerful image processing capabilities has emerged on mobile phones. One such application
domain is computer vision, which includes, e.g., feature detection, object detection and
recognition, and pattern matching.

Many feature detection algorithms and feature descriptors have been proposed, e.g.,
SIFT~\cite{lowe_object_1999}, SURF~\cite{bay_surf_2006,bay_surf_2008}, BRIEF~\cite{calonder_brief_2010},
BRISK~\cite{leutenegger_brisk_2011}, and ORB~\cite{rublee_orb_2011}.
Further, work have been done on developing such algorithms for GPUs, e.g., SIFT on desktop GPUs using
CUDA~\cite{acharya_sift_gpu_2013,yonglong_sift_gpu_2013}. For mobile GPUs, attempts have been done using
OpenGL ES 2.0~\cite{rister_sift_gpu,kayombya_sift_gpu_2010}. However, evaluation was only done using very small
images in~\cite{rister_sift_gpu} (320x240 pixels), while no evaluation was done in~\cite{kayombya_sift_gpu_2010}.
In~\cite{danielsson2016icpram}, we presented a novel feature detection/description algorithm targeting
mobile embedded devices, called Harris-Hessian/FREAK, based on a Harris-Hessian feature detector~\cite{xie_gpu-based_2010}
and a FREAK feature descriptor~\cite{alahi_freak_2012}.

The main questions addressed in this study are:
(i) How has embedded GPUs evolved the past two years, from the perspective of running a state-of-the-art feature detection algorithm?
(ii) How are the temperature and frequency behavior of the mobile GPUs when running such algorithms?


In this study, we have evaluated two generations of embedded GPUs, i.e., the Adreno 330 (in the Sony Zperia Z3) and the Adreno 530 (in the Sony Xperia XZ), when running a Harris-Hessian/FREAK feature detection algorithm. Our evaluation shows that the performance has increased a factor of ten over two generations, mainly due to more GPU cores and support for larger work-group sizes.
Further, the newer GPU was much more performance sensitive to the work-group size. Finally, we have observed that the GPUs can run at their maximum clock frequencies for long periods of time, without any thermal problems or need to reduce the clock frequency.

\section{Background and Related Work}
\label{sec:background}
Computer vision is a wide field with applications including, e.g., object recognition, image restoration and scene reconstruction.
In computer vision, \emph{feature detection} refers to methods of trying to locate arbitrary features that can afterwards be described and compared.
These features then need to be described in such a manner that the same feature in a different image can be compared and confirmed to be matching. Typically, areas around the chosen keypoint are sampled and then compiled into a vector, a so called \emph{feature descriptor}.

\subsection{Feature Detection}



Scale-Invariant Feature Transform (SIFT)~\cite{lowe_object_1999} was proposed in 1999, and has become somewhat of an industry standard. It includes both a detector and a descriptor.
The detector is based on calculating a Difference of Gaussians (DoG) with several
scale spaces.

Partially inspired by SIFT, the Speeded-Up Robust
Features (SURF)~\cite{bay_surf_2006,bay_surf_2008} detector was proposed,
which uses integral images and Hessian determinants.
SURF and SIFT are often used as base lines in evaluations of other detectors.

The detector chosen for our experiments was proposed by Xie et al. in~\cite{xie_gpu-based_2010} and
is inspired by Mikolajczyk and Schmid~\cite{mikolajczyk_scale_2004}, particularly their use of a multi-scale Harris
operator. However,
instead of increasing the scale incrementally, they
examined a large set of pictures to determine which scales should be evaluated so that as
many features as possible only are discovered in one scale each.
Then, weak corners are culled using the Hessian determinant. As the fundamental
operators are the Harris operator and the Hessian determinant, it
is called the "Harris-Hessian detector".


\subsection{Feature Description}


SIFT, SURF, and many other descriptors use strategies that are variations of histograms of gradients (HOG). The area around each keypoint in an image is divided into a grid with sub-cells. For each sub-cell, a gradient is computed. Then, a histogram of the gradients' rotations and orientations is made for each cell. These histogram then make up the descriptor. SURF, while based on the same principle, uses
Haar wavelets instead of gradients. The resulting 
descriptor vectors of a high dimension (usually $\geqslant$128) which can be compared
using, e.g., Euclidean distance.

Calonder et al. proposed a new type of descriptor called Binary Robust
Independent Elementary Features (BRIEF)~\cite{calonder_brief_2010}. Instead of using HOGs, BRIEF
samples a pair of
points at a time around the keypoint, then compares their respective intensities. The
result is a number of ones and zeros that are concatenated into a string, i.e., forming a "binary descriptor". They do not propose
a single sampling pattern, rather they consider five different ones. The resulting
descriptor is nevertheless a binary string. The benefit of binary descriptors is mainly
that they are computationally cheap, as well as suitable for comparison using Hamming
distance~\cite{hamming_error_1950},
which can be implemented effectively using the XOR operation.

Further work into improving the sampling pattern of a binary descriptor has been made,
most notably Oriented FAST and Rotated BRIEF (ORB)~\cite{rublee_orb_2011}, Binary Robust
Invariant Scalable Keypoints (BRISK)~\cite{leutenegger_brisk_2011}, and Fast Retina Keypoint
(FREAK)~\cite{alahi_freak_2012}.

The descriptor we use in this paper is FREAK~\cite{alahi_freak_2012}, where machine
learning is used to find a sampling pattern that aims to minimize the number of comparisons needed.
FREAK generates a hierarchical descriptor allowing early out comparisons.
As FREAK significantly reduces the number of necessary compare operations, it is suitable for
mobile platforms with low compute power.

\section{Harris-Hessian/FREAK}

We use the Harris-Hessian/FREAK algorithm~\cite{danielsson2016icpram}, based on a combination of the Harris-Hessian detector~\cite{xie_gpu-based_2010} and the FREAK binary descriptor~\cite{alahi_freak_2012}, as a representative feature detection algorithm targeting mobile devices.

\subsection{The Harris-Hessian Detector}
The Harris-Hessian detector was proposed by Xie et al.~\cite{xie_gpu-based_2010} and is essentially a variation of
the Harris-Affine detector combined with a use of the Hessian determinant to cull away "bad" keypoints.
The detector consists of two steps: Discovering Harris corners~\cite{harris_combined_1988} using the
Harris-affine-like~\cite{mikolajczyk_scale_2004} detector on nine pre-selected scales as well as two additional scales
surrounding the most populated one, then culling weak points using a measure derived
from the Hessian determinant.

The Harris step finds Harris corners by applying a Gaussian filter at gradually larger $\sigma$, then
reexamines the scales around the $\sigma$ where the largest number of corners
were found. This $\sigma$ is said to be the characteristic scale of the
image. After all the
scales have been explored, the resulting corners make up the scale space, $S$.

In the Hessian step, the Hessian determinant for each discovered corner
in $S$ is evaluated in all scales. If the determinant reaches a local maximum
at $\sigma_{i}$ compared to the neighboring scales $\sigma_{i-1}$ and
$\sigma_{i+1}$ and is larger than a threshold $T$, it qualifies as a
keypoint of scale $\sigma_{i}$. Otherwise, it is discarded.
The purpose of the Hessian step is to both reduce false detection and confirm
the scales of the keypoints.

\subsection{FREAK}
FREAK (Fast Retina Keypoint) is a so called ``binary'' descriptor,
since its information is represented as a bit string.
Alahi et al.~\cite{alahi_freak_2012} propose a circular sampling pattern of
overlapping areas inspired by the human retina. They then---optionally---define
45 pairs using these areas and examines their gradients, to estimate the orientation
of the keypoint. With the new orientation, the pattern is rotated accordingly and
areas are re-sampled. They use machine learning to establish which pairs of areas
result in the highest performance for the descriptor bit string. The sampling pairs
are sorted into four cascades with 128 pairs each, starting with coarse (faraway)
areas and successively becoming finer and finer.
This finally results in a bit string with 512 elements.

\section{Implementation}


A more detailed description of our implementation is found in~\cite{danielsson2016icpram},
so we only provide a high-level description here.
Our implementation is written in standard C99 and OpenCL 1.1~\cite{munshi_opencl_2011},
and compiled, built and installed using the Android SDK and NDK toolsets.
Additionally, we utilize
\texttt{stbi\_image\footnote{Sean Barret,~\url{http://nothings.org/}}} and
\texttt{lodepng\footnote{Lode Vandevenne,~\url{http://lodev.org/lodepng/}}} for
image decoding/encoding, \texttt{ieeehalfprecision\footnote{Developed by
James Tursa.}} for half-float encoding, and Android Java to create an application
wrapper.

All calculations are done in a raster data format, and we maintain the same resolution as the original image.
We convert the image to grey scale as the algorithms
do not account for color. 
We normalize and represent scalar pixel values as
floating point values in the range of 0.0 to 1.0.

\subsection{Algorithm Overview}

\begin{figure}[ht]
    \centering
    \includegraphics[width=1.0\columnwidth]{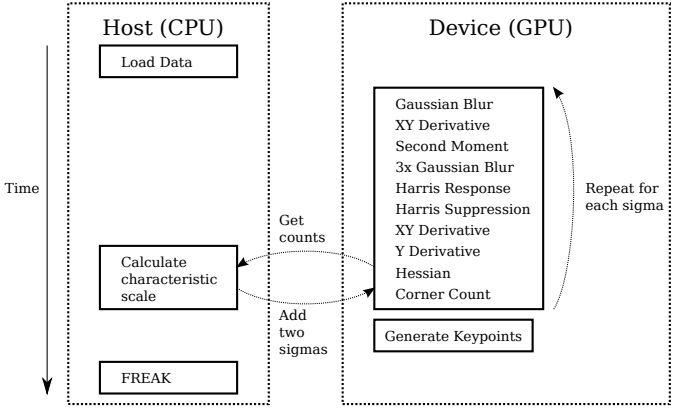}
    \caption{Visual
    representation of the algorithm. On the left side is the host CPU with
    initialization of data, keypoint counts, and execution of FREAK. On
    the right is the twelve executional kernel calls to perform Harris-Hessian for
    a given scale and finally the keypoint generation kernel call which gathers the
    resulting data. Execution order is from top to bottom.}
    \label{fig:impl}
\end{figure}

The program is executed in a number of steps, see Fig.~\ref{fig:impl},
starting with setting up buffers, loading image data, and decoding it
into a raw raster format. The image is transferred to the device before execution
of Harris-Hessian and desaturation is performed on the GPU as a separate
step.

\subsection{Harris-Hessian}

\begin{figure}[!tbh]
    \centering
    \includegraphics[width=1.0\columnwidth]{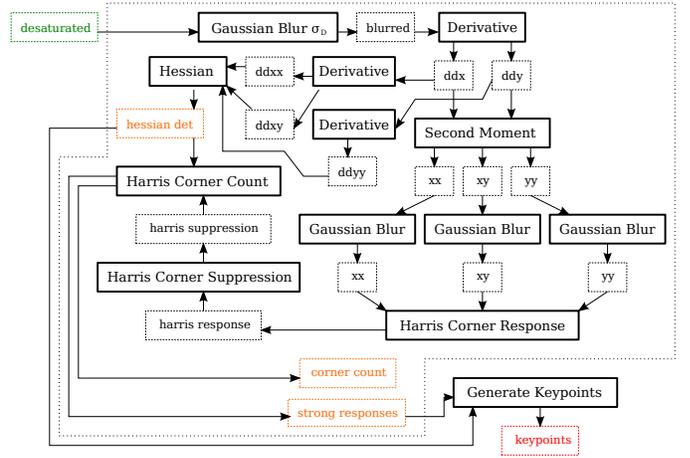}
    \caption{Data flow in Harris-Hessian. Solid boxes indicate kernel executions
and the dotted boxes are buffers or data. Green boxes are input and orange are
the resulting output for a given sigma. Red boxes are the results sent to the
descriptor. The larger dotted border indicates sigma iteration, anything within
this border is executed for each sigma.}
    \label{fig:dataflow_hh}
\end{figure}

The implementation is split into two main parts: the Harris-Hessian detector
and the FREAK descriptor. Fig.~\ref{fig:dataflow_hh} shows an overview of our implementation of the Harris-Hessian detector. Our implementation is targeted for GPU execution, and based on the description in~\cite{xie_efficient_2011}.
Harris-Hessian is first executed for the sigmas $0.7, 2, 4, 6, 8, 12, 16, 20,
24$. For each sigma, the number of corners are counted and the sums are transferred
to the CPU, which then calculates the characteristic sigma. After the
characteristic sigma $\sigma_c$ is found, we run the Harris-Hessian two more
times for $\frac{\sigma_c}{\sqrt{2}}$ and $\sigma_c\cdot\sqrt{2}$.


A majority of GPU execution is spent in the Gaussian blur kernels.
A $\sigma = 20$ results in a 121 elements wide filter, i.e.,
$121 * 2$ global memory accesses per task which is significant compared to all
other kernels. 
Therefore, we use prefetching in the Gaussian kernel, i.e.,
preloading the global memory into local work-group shared memory. For a work-group
(8 by 4 tasks) running the x axis Gaussian kernel, we
perform a global to local memory fetch of $(60 + 8 + 60) * 4$ elements
and then access the shared local memory from each task.

After running Harris-Hessian, we generate a list of keypoints
containing the sigma and coordinates. The keypoints are passed to the
FREAK algorithm together with the source image. FREAK then calculates a 512-bit descriptor
for each keypoint, which is written to an external file.

\subsection{FREAK}

The FREAK implementation runs on the host CPU and is based on the
implementation in~\cite{alahi_freak_2012}\footnote{Source can be
found at \url{https://github.com/kikohs/freak}}.
The main differences in our implementation
compared to the origial~\cite{alahi_freak_2012} are:
we do not utilize SIMD instructions,
we always take rotational or scale invariance into account, and
we only use a generated and hard-coded sampling pattern.


\section{Experimental Methodology}

Our experiment and measurements were conducted on a Sony Xperia
Z3~\cite{Z3} and on a Sony Xperia XZ~\cite{XZ}. Table~\ref{tab:hw} summarizes the
main hardware characteristics of the two phones.
The presented execution times are the mean of ten runs.
The CPU and GPU temperature and frequency measurements were done using
internal probes on the chipset.
When running the temperature tests the phone was placed on a table, standing
up with the back leaning towards a surface touching a small part of the phone.
The room's temperature was around \SI{20}{\celsius}.

\begin{table}[!tbh]
\caption{Hardware characteristics of Sony Xperia Z3 and Sony Xperia XZ.}
\label{tab:hw}
\vspace*{-3ex}
\begin{center}
\begin{tabular}{l @{\hspace{2em}} c @{\hspace{2em}} c}
  & Xperia Z3~\cite{Z3} & Xperia XZ~\cite{XZ} \\ \hline
  Release date & Sep./Oct. 2014 & Oct. 2016 \\
  Chipset & Snapdragon 801 & Snapdragon 820 \\
  CPU & Krait 400 & Kryo \\
  CPU cores & 4 & 4 \\
  CPU frequency & 2.5 GHz & 2.15 GHz \\
  GPU & Adreno 330 & Adreno 530 \\
  GPU cores & 128 & 256 \\
  GPU frequency & 450/550/578 MHz & 510/624/650 MHz \\
  Main memory & 3 GB & 3 GB \\
  Flash memory & 16 GB & 32 GB \\
\end{tabular}
\end{center}
\end{table}

\begin{figure}[!ht]
    \centering
    \includegraphics[width=0.8\columnwidth]{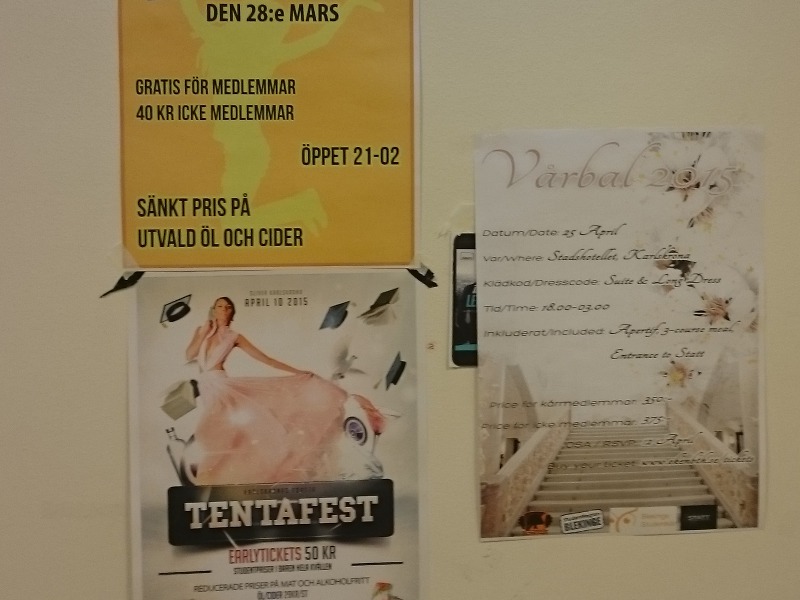}
    \caption{Our test image, 800x600 pixels, featuring a series of posters.}
    \label{img:test}
\end{figure}

As input in our experiments, we use the image shown in Fig.~\ref{img:test}.
The image content has little effect on Harris-Hessian algorithms. However, it has
an impact on FREAK, since different images have different
numbers of keypoints and FREAK scales linearly with the number of descriptors.
We have not set any limitations on the number of
descriptors encoded, which is relevant in a final
implementation as it affects both the execution time and the storage
requirements for the descriptor.

\section{Experimental Results}
\label{sec:results}

\subsection{Kernel Execution Times}
\label{sec:exetimes}

\begin{figure}[!tbh] 
    \centering
    \includegraphics[width=0.5\columnwidth,angle=-90]{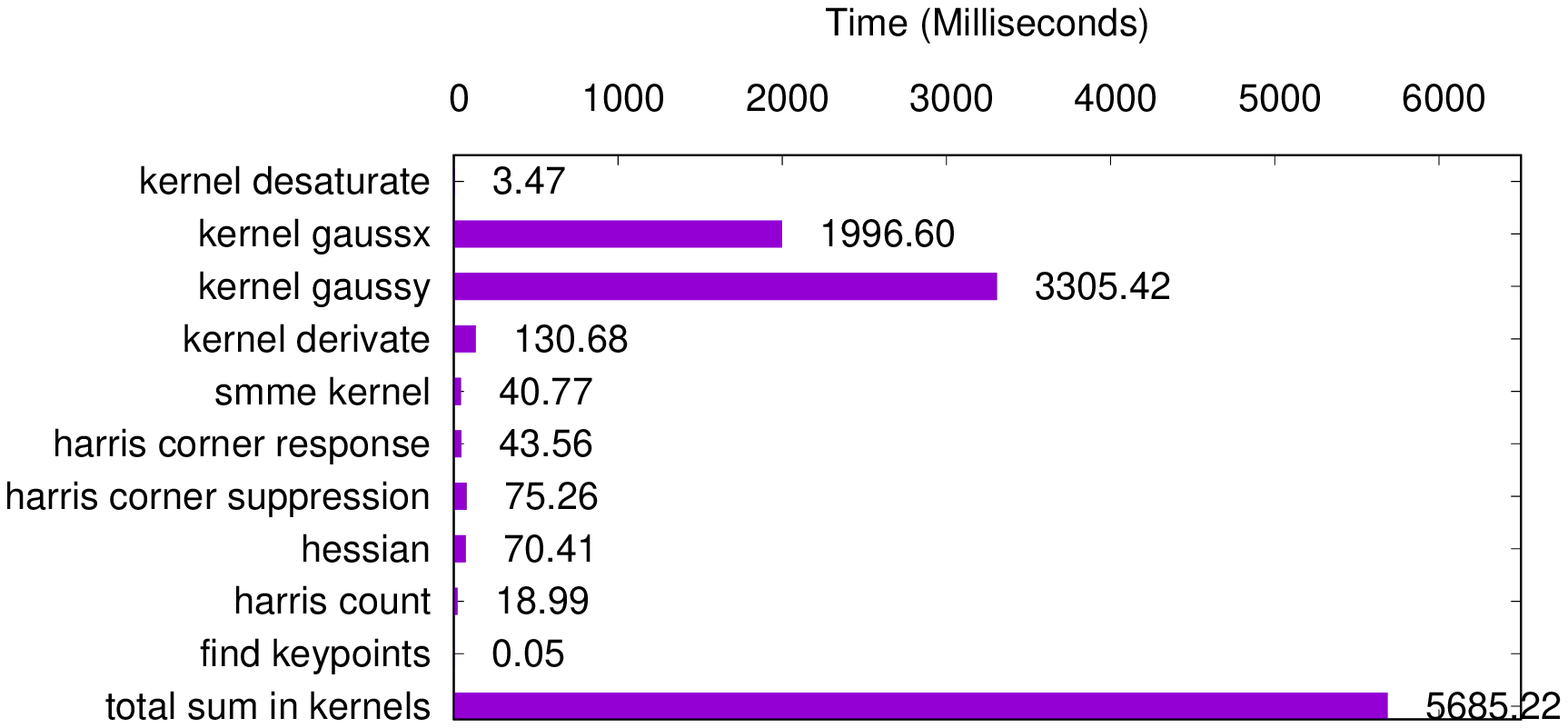}
    \includegraphics[width=0.5\columnwidth,angle=-90]{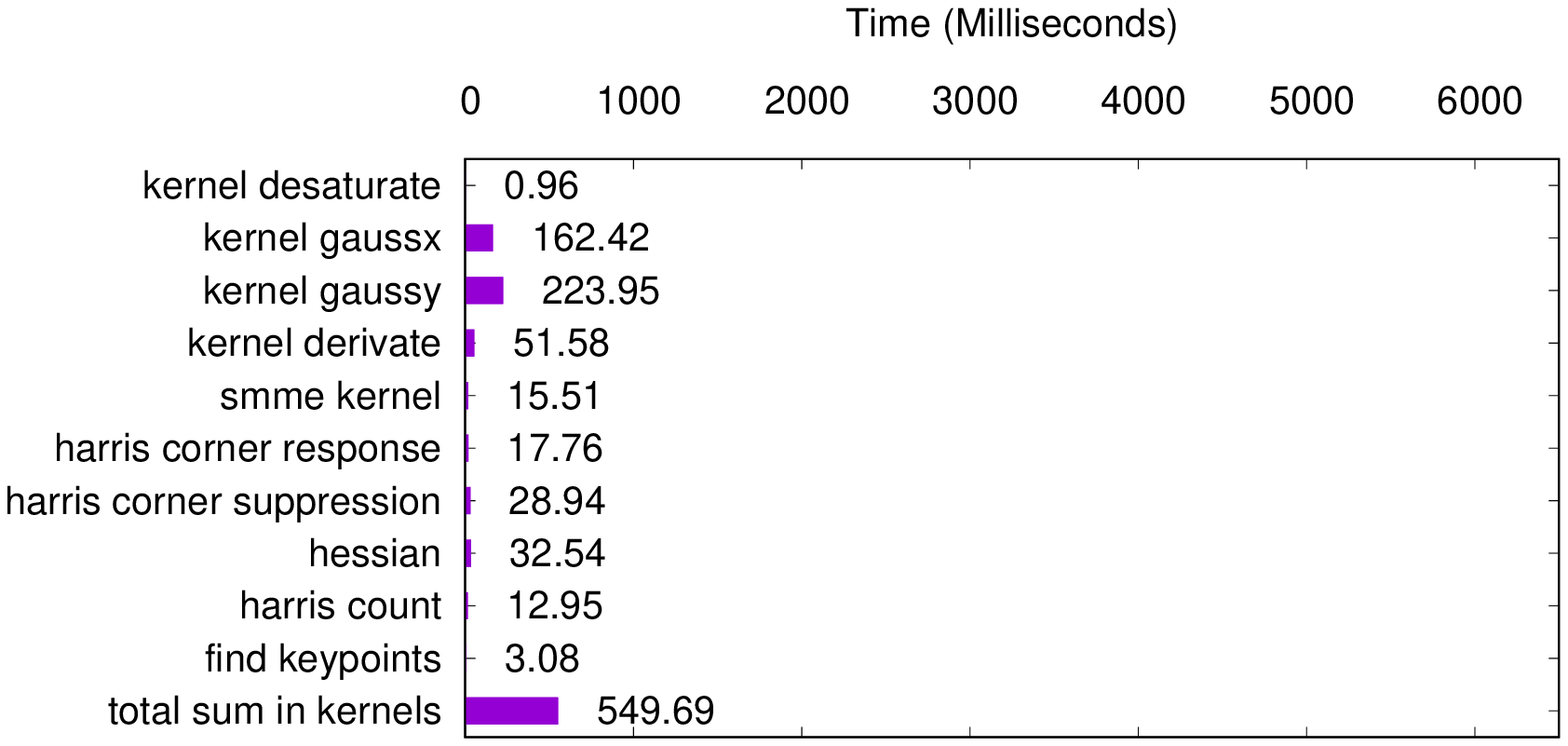}
    \caption{Execution times (mean of 10 runs) on Xperia Z3 (upper) and Xperia XZ (lower) for the individual kernels.}
    \label{fig:phone_performance}
\end{figure}

In Fig.~\ref{fig:phone_performance}, we present the execution times for the
different GPU kernels running on Xperia Z3 (upper) and Xperia XZ (lower).
We present the mean time of ten executions, however, the times varied very little between the runs.

Our main observation in Fig.~\ref{fig:phone_performance} is that the total time for the GPU kernels was reduced by a factor of ten, i.e., from almost 5700 ms to 550 ms, when moving from the Z3 to the XZ. Clearly, the reason is not only twice as many GPU cores on XZ, see Table~\ref{tab:hw}. We identified that one main reason is that the GPU in the XZ supports larger work-group sizes (up to 1024 vs. 256). Therefore, we evaluated how various workgroup sizes impact the performance.

\subsection{Effect of Various Workgroup Sizes}
\label{sec:workgroup}

As we saw in the previous section, the two kernels \texttt{gaussx} and \texttt{gaussy} contribute most to the execution time. Therefore, we have focused on them when we evaluated the effects of various workgroup sizes. We varied the work-group sizes between $2\times2$ up to $32\times8$ on the Xperia Z3 and between $2\times2$ up to $128\times8$ and $256\times4$ on the Xperia XZ when executing the Gaussian blur kernels.

\begin{table}[tbh]
\centering
\caption{Best and worst execution times (in ms) for different work-group sizes for the GaussX and GaussY kernels, \newline
along with the work-group sizes.}
\label{tab:workgroups}
\begin{tabular}{l @{\hspace{2em}} c @{\hspace{2em}} c}
GaussX & Best & Worst \\ \hline
Xperia Z3 & 639 ms (32x8)  & 4560 ms  (2x2) \\
Xperia XZ & 162 ms (128x8) & 12833 ms (2x2) \vspace*{2ex} \\

GaussY & Best & Worst \\ \hline
Xperia Z3 & 676 ms (8x32)  & 6234 ms  (2x2) \\
Xperia XZ & 224 ms (2x256) & 15846 ms (2x2) \\
\end{tabular}
\end{table}

\begin{figure*}[!ht]
    \centering
    \includegraphics[width=0.49\textwidth]{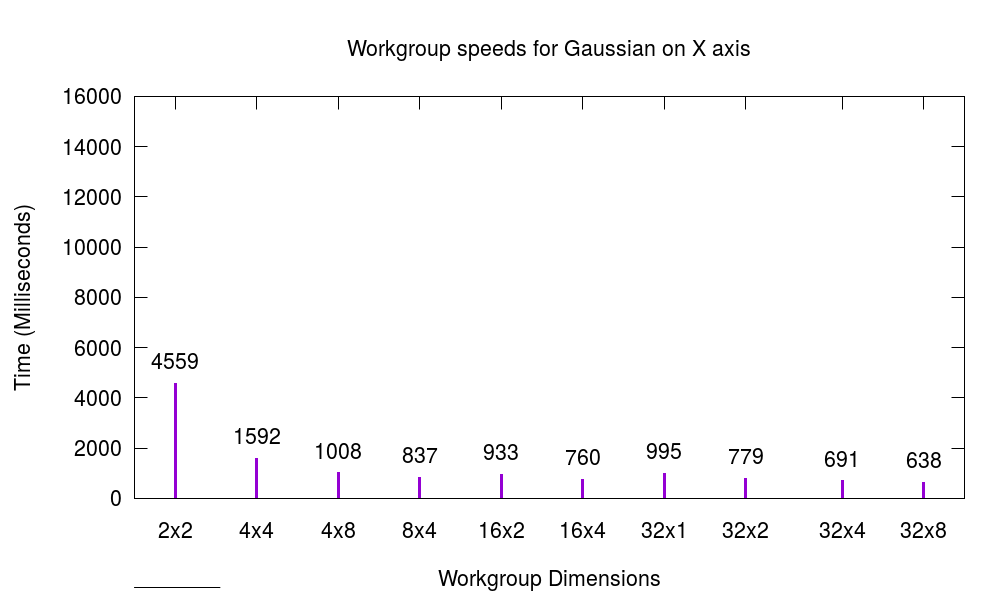}
    \includegraphics[width=0.49\textwidth]{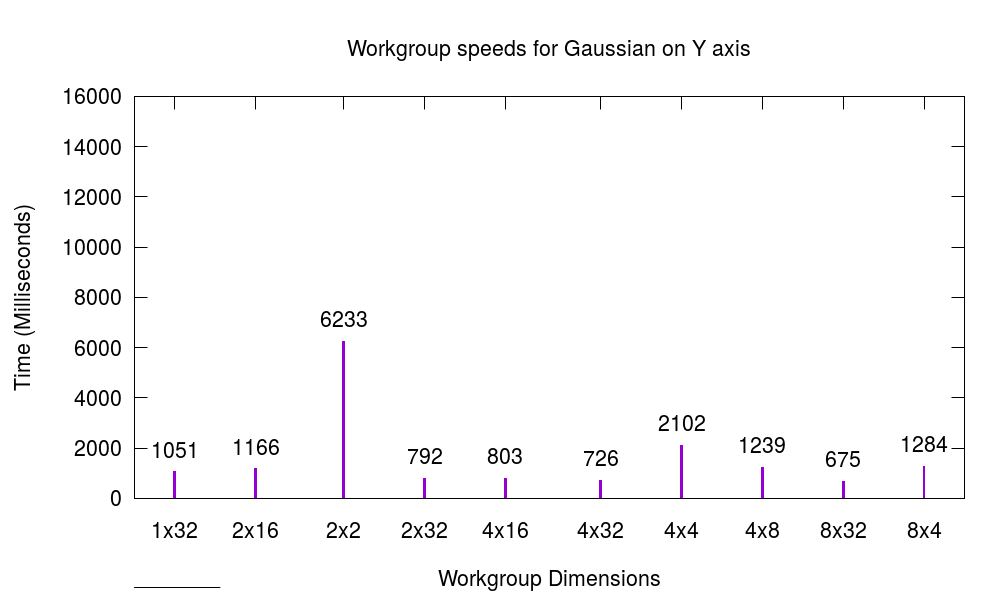}
    \caption{Execution times on Xperia Z3 for various workgroup sizes for the \texttt{GaussX} (left) and \texttt{GaussY} (right) kernels.}
    \label{fig:z3workgroup}
\end{figure*}

\begin{figure*}[!ht]
    \centering
    \includegraphics[width=0.94\textwidth]{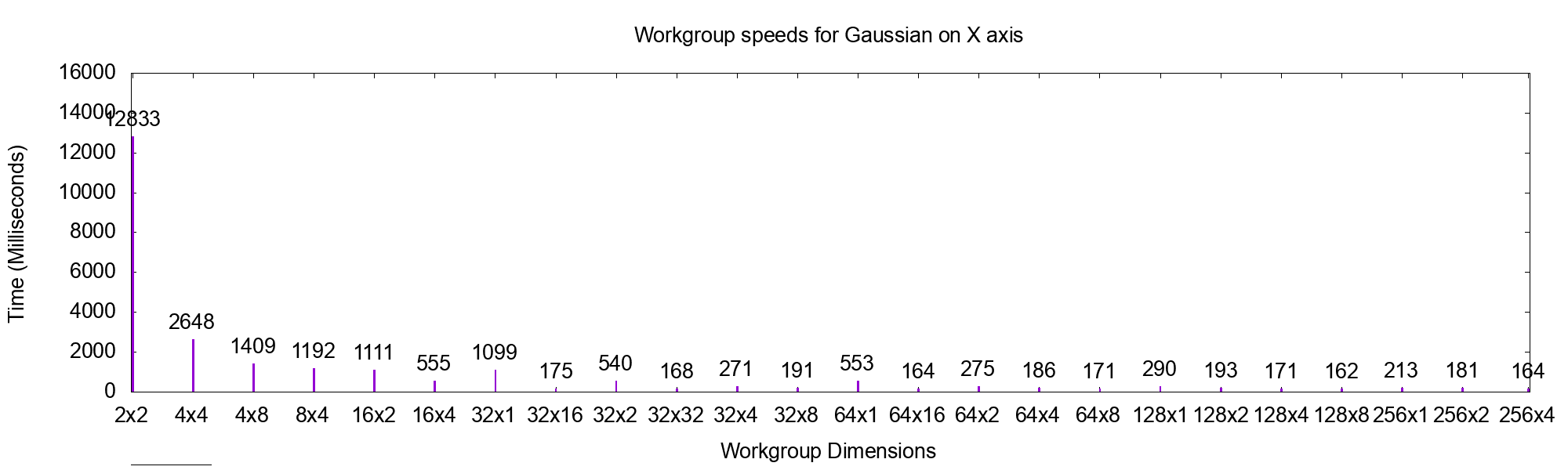} \\
    \includegraphics[width=0.94\textwidth]{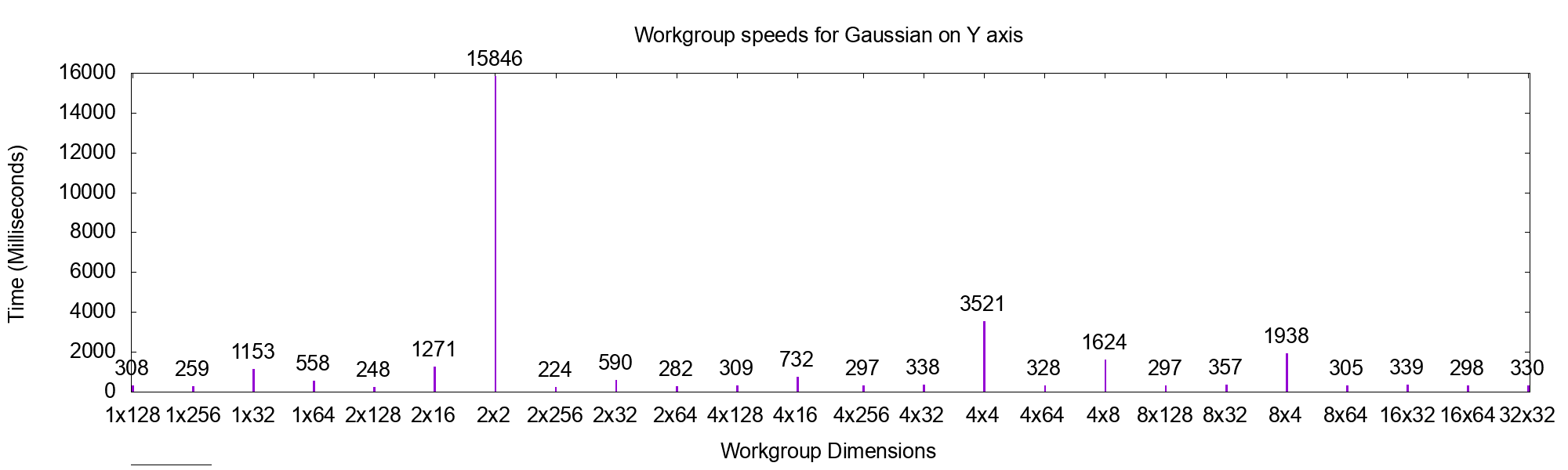}
    \caption{Execution times on Xperia XZ for various workgroup sizes for the \texttt{GaussX} (upper) and \texttt{GaussY} (lower) kernels.}
    \label{fig:xzworkgroup}
\end{figure*}

The execution times vary significantly, as shown in Figure~\ref{fig:z3workgroup} and Figure~\ref{fig:xzworkgroup}. In Table~\ref{tab:workgroups} we summarize the best case and worst case on each of the phones. On the Z3 we observed variations of up to a factor of ten, but on the XZ the variation was even larger. The worst kernel execution time was almost 80 times slower than the best on the XZ. Therefore, we conclude that a proper selection of the work-group size has a significant impact on the GPU execution time on the XZ (Adreno 530 GPU).

\subsection{Temperature Effects}
\label{sec:temperature}

\begin{figure*}[!ht]
    \centering
    \includegraphics[width=0.9\columnwidth]{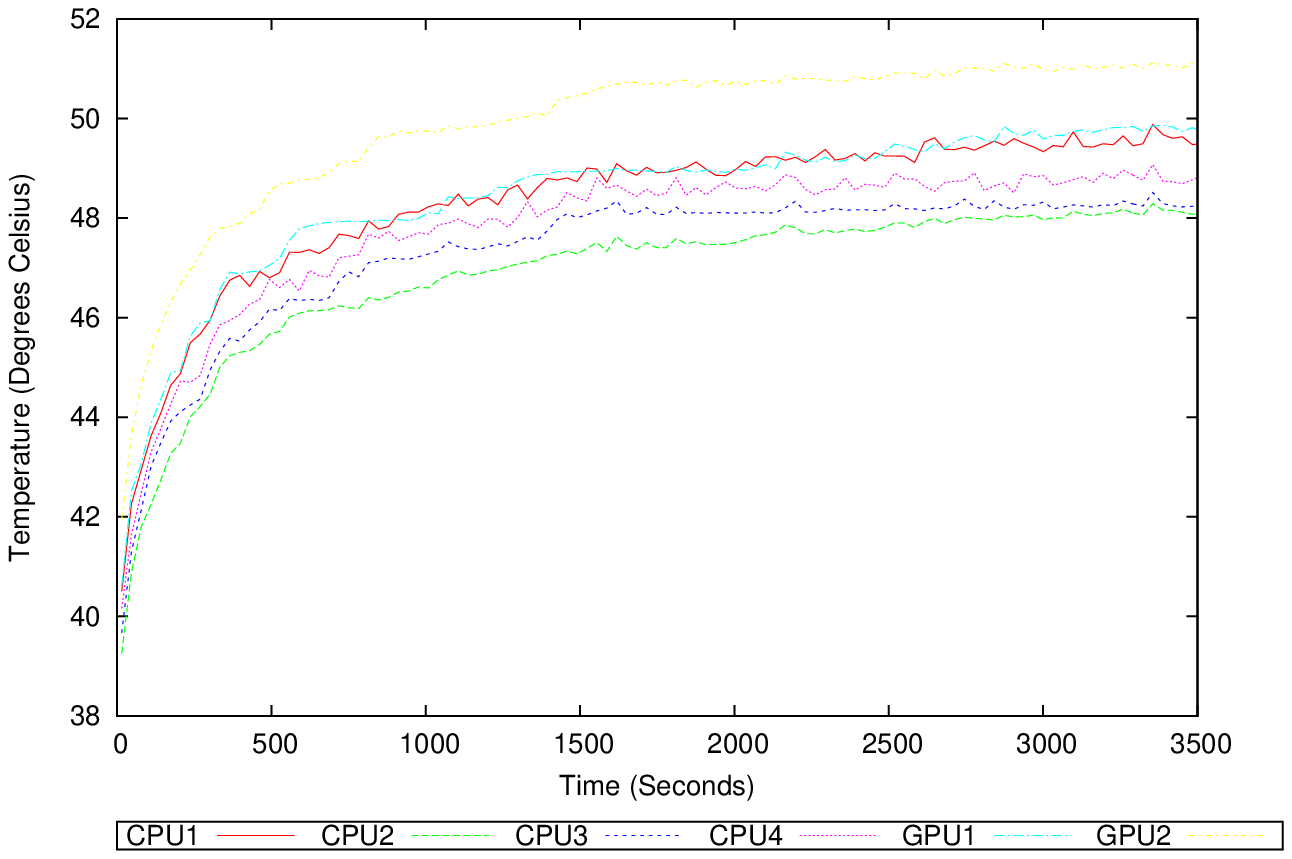}
    \hspace*{2em} 
    \includegraphics[width=0.9\columnwidth]{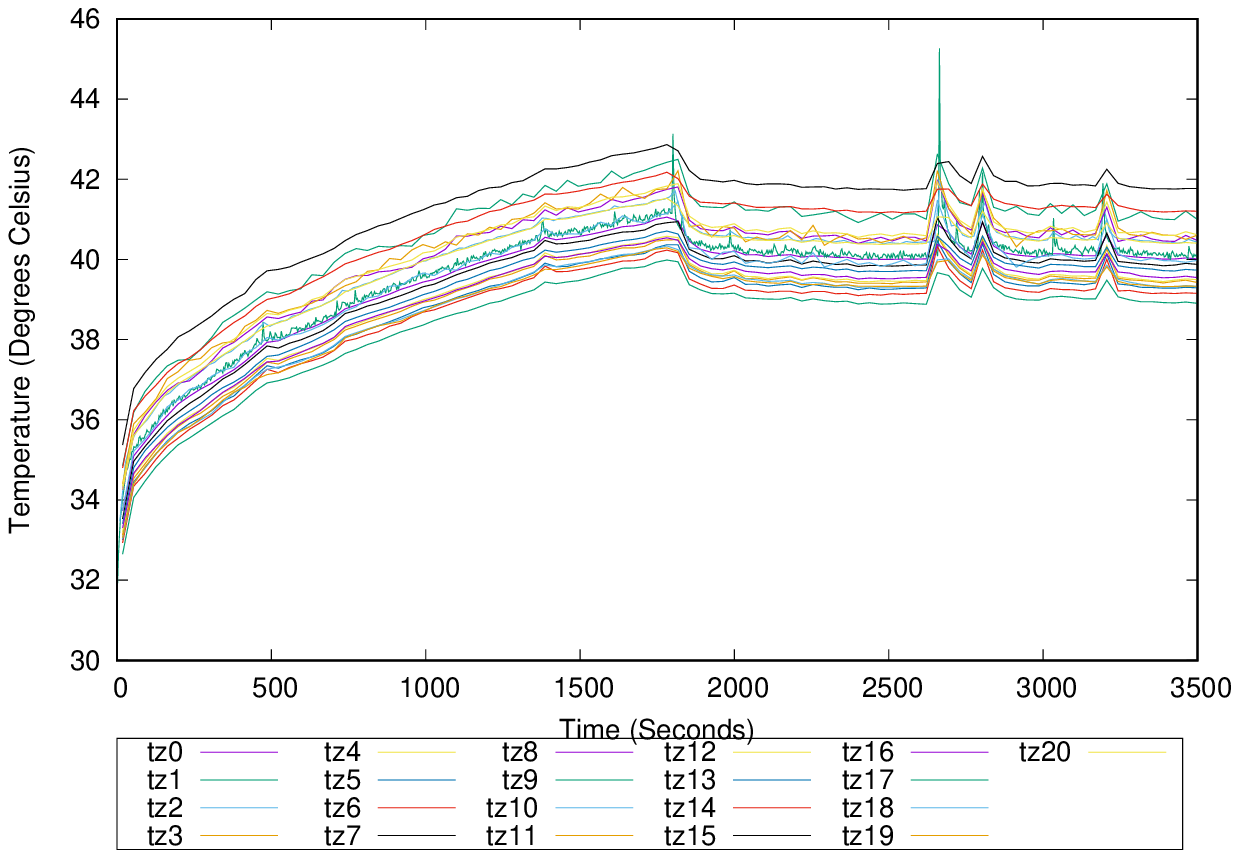}
    \vspace*{-2ex}
    \caption{Temperature during 3500 seconds, from start of idle phone, on
    Sony Xperia Z3 (left) and Sony Xperia XZ (right).
    }
    \label{fig:heat}
\end{figure*}

The second aspect that we evaluated is the operational temperature of the phones when running a performance demanding computer vision algorithm.
Our results in Fig.~\ref{fig:heat} indicate that neither the Xperia Z3 nor the
Xperia XZ have any temperature issues when running the Harris-Hessian/FREAK application.

When the program starts, the phones have been idle for a significant period of
time, and we see that they have a temperature of approximately \SI{38}{\celsius} (Z3) and \SI{35}{\celsius} (XZ).
After running the program for roughly 30 minutes, we see that the Z3 has reached a stable temperature
zone around \SI{50}{\celsius} for the GPU sensors.
On the XZ, we have not been able to map the different temperature sensors (tz0-tz20) to specific parts of the chip set,
but we can conclude that the XZ has a stable working temperature between \SI{38}{\celsius} to \SI{42}{\celsius}.

\begin{figure}[!ht]
    \centering
    \includegraphics[width=1.0\columnwidth]{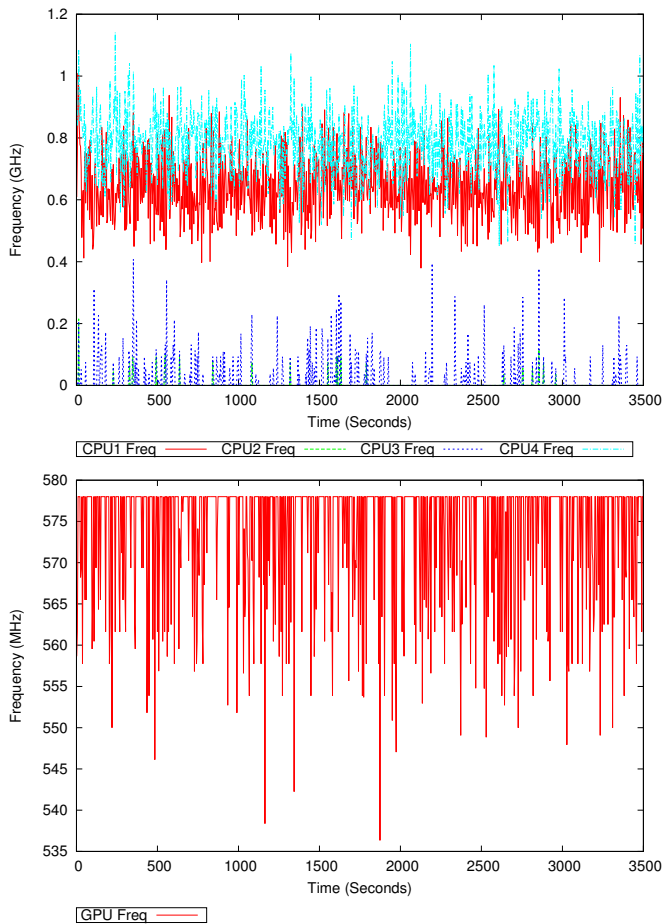}
    \includegraphics[width=1.0\columnwidth]{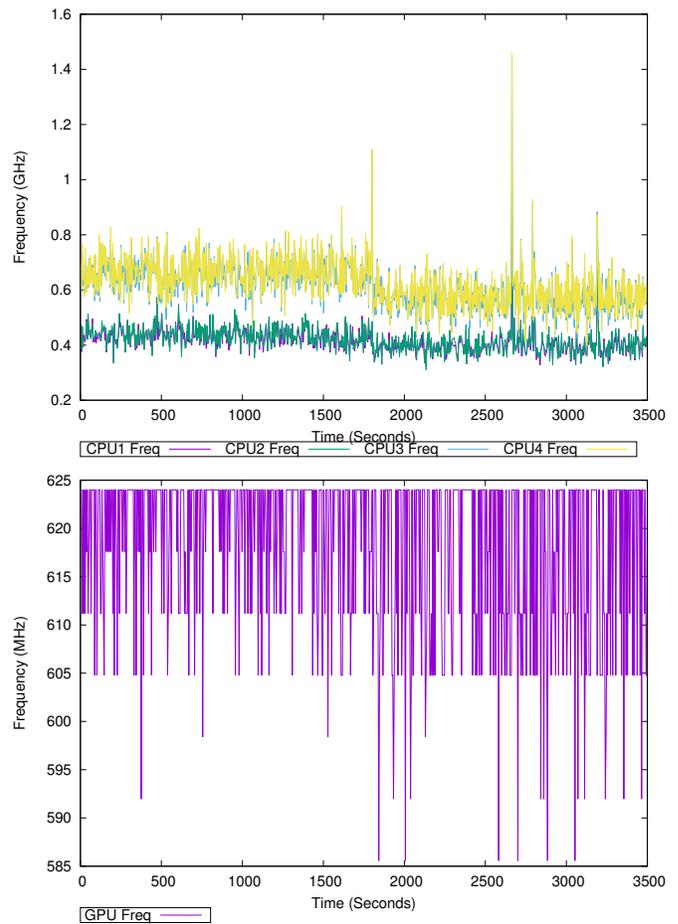}
    \caption{Clock frequency of the CPUs and the GPU of the Xperia Z3 during the heat stress test.
    In the graphs we see that CPU 1 and CPU 4 both are active while CPU 2 and 3 appear inactive.
    The GPU is mainly running at maximum speed with occasional short dips.}
    \label{fig:z3_freq}
\end{figure}

\begin{figure}[!ht]
    \centering
    \includegraphics[width=1.0\columnwidth]{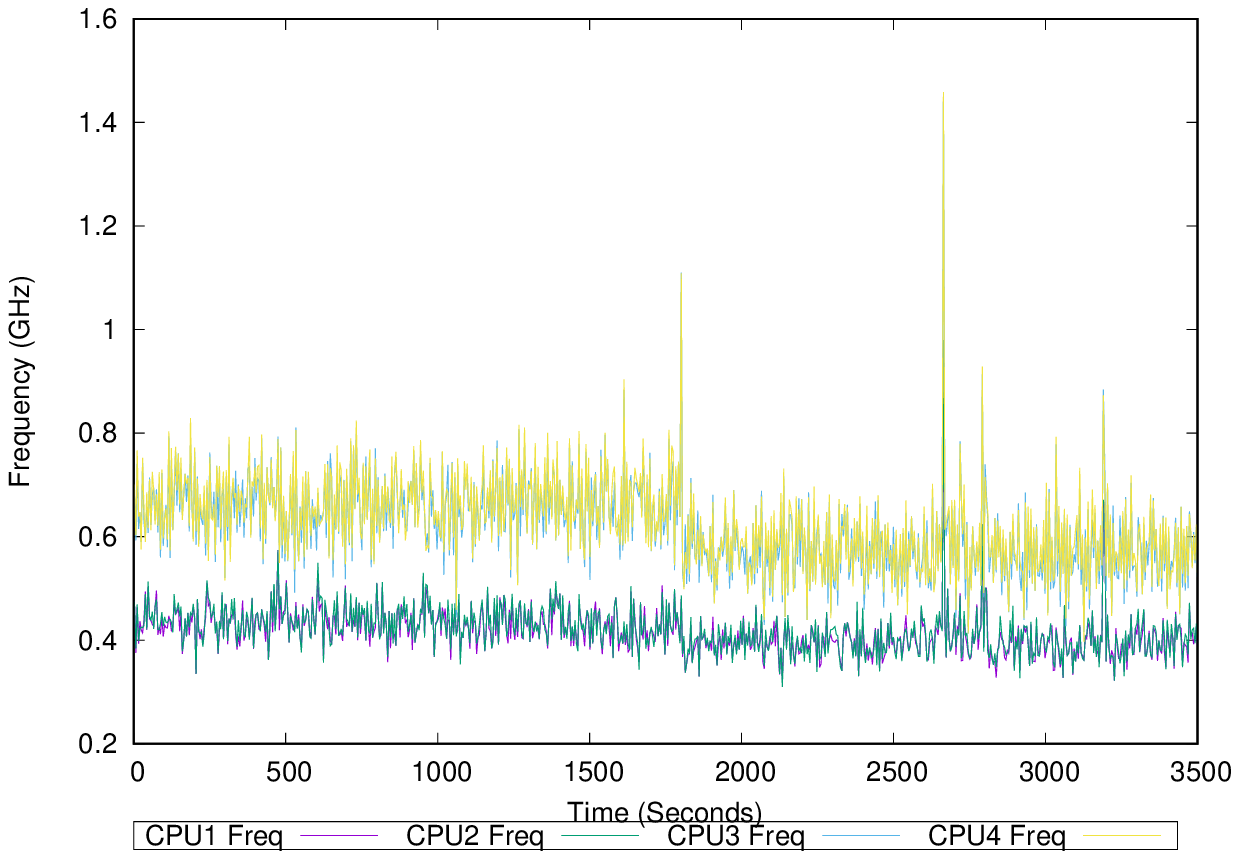}
    \includegraphics[width=1.0\columnwidth]{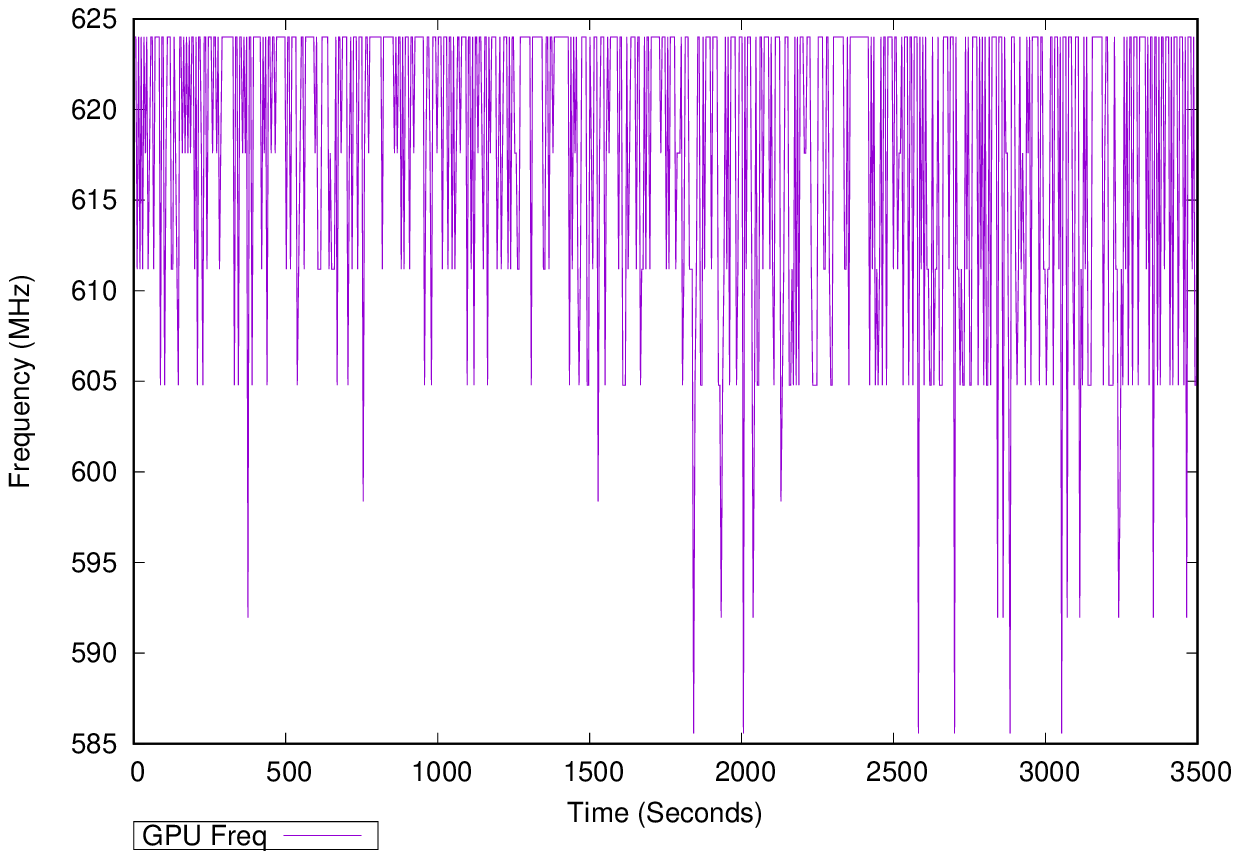}
    \caption{Clock frequency of the CPUs and the GPU of the Xperia XZ during the heat stress test.
    In the graphs we see that CPU3 and CPU4 both are active while CPU1 and CPU2 have less work to do.
    The GPU is mainly running at maximum speed with occasional short dips.}
    \label{fig:xz_freq}
\end{figure}

On the XZ, we can also observe that after approximately 1800 seconds, the temperature drops approximately \SI{1}{\celsius} on the XZ.
This can be correlated to the frequency measurements in Fig~\ref{fig:xz_freq}. After approximately 1800 s we observe that the working
frequency for CPU3 and CPU4 drop $\approx100$ MHz (from $\approx700$ MHz to $\approx600$ MHz).
A general observation from the frequency measurements in Fig.~\ref{fig:z3_freq} and Fig.~\ref{fig:xz_freq} is
that only 2 CPU cores appear active in both the Z3 and the XZ, while the GPU runs at max frequency for the
majority of the program execution. The GPU frequency drops mainly when the program is running the
FREAK algorithm, which is exclusively on the CPU.

\if
We have been informed that the 50 degree region is when the phone is set to mitigate temperature
by selectively lowering the clock or temporarily disabling devices. However, the data in Fig.~\ref{z3_freq} does not indicate
that this is in fact happening, as the frequency patterns and execution times are largely unchanged
for the duration of the program. The GPU frequency drops that are apparent could be assigned to the
switch from Harris-Hessian to FREAK execution. The apparent peaks in temperature align with
execution of FREAK which increases the CPU load. This peak does reach values over 50 degrees which might
make the mitigation kick in, but if that is the case, the unchanging execution time tells us that it is
happening for all runs.
\fi

\section{Conclusion}

In this paper we have studied two generations of embedded GPUs when running a performance
demanding computer vision algorithm. Our study indicates that the performance has
increased a factor of ten over two generations, mainly due to more GPU cores and
support for larger work-group sizes. Further, the newer GPU was much more performance
sensitive to the work-group size.

We have observed that the GPUs can run at their maximum clock frequencies for long
periods of time, without any thermal problems or need to reduce the clock frequency.
In contrast, the CPU frequencies were decreased to reduce the working temperature.

\section*{Acknowledgments}
This work was partly funded by the Industrial Excellence Center "EASE - Embedded Applications Software Engineering", (http://ease.cs.lth.se), and the "Scalable resource-efficient systems for big data analytics" project funded by the Knowledge Foundation (grant: 20140032) in Sweden.

\newpage
\bibliographystyle{plain}
\bibliography{esl_references}



\end{document}